\documentclass[runningheads]{llncs}
\usepackage{graphicx}
\usepackage[hyphens]{url}
\usepackage{hyperref}
\hypersetup{%
    pdfborder = {0 0 0}
}
\hypersetup{colorlinks=false,breaklinks=true}
\begin{document}
\title{Perlustration on Mobile Forensics Tools}
\author{Utkarsha Shukla\and
Bishwas Mandal \and
K.V.D Kiran}

\authorrunning{U. Shukla, B. Mandal et al.}
\institute{Koneru Lakshmaiah Education Foundation, Vaddeswaram, AP, India \\
\email{\{utkar24, bishwasmandal246\}@gmail.com, \ kiran\_cse@kluniversity.in}}
\maketitle              % typeset the header of the contribution
\begin{abstract}
Nowadays many people store more information in cellphones rather than do on their computer which leads to the increase of crimes taking place in mobile. This is also because of extensive use of mobile devices. People refer to store more information in mobile than in computer. Therefore,scope of forensics in mobile has increased in last few years as compared to computers. In this paper , we have carried out a survey on tools and techniques which are in practice for mobile forensics. The survey is the amalgamation of the in practice technical methods and discussion of some of the tools for specific operating systems. Some challenges faced by practitioners while performing forensics has also been enlisted.

\end{abstract}
\section{Introduction}
According to National Institute of Standards and Technology(NIST) digital forensics is defined as “an applied science to identify an incident, collection, examination and analysis of evidence data. In digital forensics, it is mandatory to maintain the integrity of the information and strict chain of custody for the data. Many other researchers have defined digital forensics as the procedure of examining a computer system in order to determine the potential legal evidence~\cite{7214075}. Digital Forensics has many applications. The most common is to support a hypothesis before criminal or civil courts where evidence is collected support or oppose the hypothesis given by court~\cite{numb2}. The goal of this process is to integrate the evidence in its original form by collecting, identifying and validating digital information. Digital forensics is becoming important because our society is becoming more independent on various communication tools and technologies.
Digital forensics is subdivided into subsections such as Computer forensics, Network Forensics, Malware Forensics, Mobile Forensics, Memory Forensics and OS Forensics~\cite{7973613}. Mobile Forensics aims at the retrieval or gathering of data and evidence form mobile phone and similar devices used in daily life. Due to constant upgrades, changes and new additions of mobile phones, there has been lack of forensic tools for retrieval that is compatible with today’s uprising mobile technologies~\cite{Sai2015TheFP}. In this paper, we’ll discuss about the Mobile Forensics and the tools involved in performing mobile forensics activities.

\section{Mobile Forensics}
Mobile forensics refers to one of the types of digital forensics which deals with the recovery of  digital evidence or data from a mobile device under suitable conditions required for forensics. The term mobile device refers to any digital device that has internal memory along with communication ability, PDA devices, GPS devices and tablet computers. Mobile forensics consists of four main steps which are illustrated below:

\paragraph{Identification:} It refers to the location of evidence. It generally deals about the preservation  of the data which makes sure that the data maintains integrity. One of the best ways to isolate a mobile phone device is by putting it in a Faraday bag which prevents the transmission of electromagnetic waves, heating and electric shock. Policy of chain of custody should be followed which deals with the details regarding the handover of evidence from one individual to other. Another important phase of identification is taking pictures of the crime scene and capturing the present state of the device.

\paragraph{Acquisition:} It is the process of acquiring the evidence by maintaining the integrity of the data.

\paragraph{Examination:} During an examination phase an investigator usually recovers the deleted or lost data in order to extract information. The type of data recovered depends on the type of investigation, which includes emails, log files, internet history or documents. Data can be recovered from accessible disk space, deleted files or from within cache files of the operating system.

\paragraph{Report:} Reporting is a comprehensive summary of the results obtained by investigators during the examination phase. It contains detailed description of each and every process (or thing) which has taken place during investigation~\cite{numb5}.

\section{Mobile Forensics Techniques}
Forensic software tools are continually developing new techniques and methods for extraction of data from several cellular devices. Some common techniques of extraction of data from cellphones are described below:

\paragraph{Manual extraction:} This technique allows investigators to extract and view data through the device’s touch screen or keypad. 

\paragraph{Logical extraction:} Here, investigators connect cellular devices through USB or Bluetooth. The tool connected through a computer sends series of commands through the computer and analysis is done accordingly.  

\paragraph{Hex Dump:} This technique is also known as physical extraction in which extraction of raw image is done in binary format from the mobile device. Boot loader is pushed into the device which instructs the device to dump its memory to the computer. Recovery of phone’s deleted files and unallocated space is also taken care in this technique. 

\paragraph{Chip-Off:} In this technique data is extracted from flash memory of the mobile. Phone’s chip is removed and its binary image is created and data is extracted from the binary image. Forensic expert should have sound knowledge for using this technique during investigation.  

\paragraph{Micro Read:} In this technique data is interpreted and viewed on memory chips. Here, the physical gates on the chips are analyzed and are converted to ASCII code. This process is time consuming and a lot of knowledge is required to use this technique~\cite{numb6}.

\section{Mobile Forensics Tools}

\subsection{MOBILedit}
MOBILedit is a digital forensic software that has been designed and developed by compelson labs to search, examine and report the data form GSM/ CDMA/ PCS cell phones. The software can be connected through cell phone by the medium of infrared port, bluetooth link, Wi-Fi , or cable interface. After the connection, the phone model is identified by various parameters such as manufacturer, model number, and serial number (IMEI) along with a corresponding picture of the phone. It performs the following operations which are outlined below:

\paragraph{Complete extraction of data from phones and sim card:} We can search or retrieve whatever data we want by applying few clicks on the tool which includes, phone books, text messages, multimedia items files, calendars  etc. We can also rectify the information about the details of operating system of the phone, firmware information, sim details and many more.

\paragraph{Support for all kinds of cell phones:} The software supports all kinds of cell phones, irrespective of the brand such as Nokia, Samsung, Xiaomi, etc  and operating system like Android and IOS.

\paragraph{Bypass the passcode on ios using the lockdown files method:} IOS data is well protected and secure but MOBILedit manages to extract the data from it. The data is extracted with the assistance of lockdown files method . The former method is generated by the system by typing the passcode. Lockdown files are imported from the system by using data acquisition, once data is acquired then there is no use of passcode to extract the data.

\paragraph{Bypass the PIN code by using SIM Cloning tool:} This features removes the necessity of PIN code from the mobile as well as the need of outdated and unreliable Faraday bags.  The feature helps to clone SIM cards, or connect multiple SIM card readers at the same time.

\paragraph{Examine the phone data without having the phone:} This feature helps in automatic backup of the data available in the phone. We have observed that nowadays automatic backup is done of our whatsapp chats , viber chats even if we are not willing to do. MOBILedit allows us to obtain backup files which contains messages, images, recordings of whatsapp or viber.

\paragraph{Providing assistance while connecting the mobile to computer:} The software contains unique driver detection which automatically downloads and installs the correct driver required for the extraction and acquisition of data from mobile to computer~\cite{numb7}

\subsection{Athena}
Athena was initially named as Radio Tactics. It has a touch screen technology because of which users are able to understand the device completely. Its security level is high and requires smart card and pin card to access it. It allows the user to think and decide on selecting proper data to retrieve and specify amount of time required to finish the selected chore. It has got the ability of backing up the data if the device is unexpectedly terminated or when it fails to load. Athena is a completely movable and vigorous with long lasting battery. 

\subsection{Lantern}
Lantern is developed by KATANA forensics It is considered as one of the most effective and   synopsis providing tool. It is a MAC based tool which is used in case of iPhone, iPod Touch and iPad. It has got a feature of keyword searching based on dynamic text and SMS database. The tool can extract more data as compared to any other tool in the market. Dynamic text database is a keying sequence for the iPhone.  An iPhone user will add this feature to the database from various applications to the device. It has the ability to show the list of all deleted SMS. One unique feature it has is that it has a voicemail riveted to it. The tool’s focus on a single device allows the investigators to validate the information more easily than they can do with any other tools. 

\subsection{SIMIS}
SIMIS is among the earliest mobile forensic tools and today it is one of the most synoptic tools available all across the globe. It is generally use to analyze SIM card of the mobile. It was engineered according to Association of Chief Police Officers, ACPO guidelines to protect the integrity of the data on the SIM. Here, the data format is UNICODE in order to display native characters. Hashing of data is done in MD5 and SHA1 format. SIMIS is the perfect tool to analyze data from SIM cards. SIM has the ability to build database with unique file reference of each SIM card which ensures integrity of this tool.

\subsection{Neutrino}
NEUTRINO is perhaps the only device which becomes interlinked with Encase V6 which helps to analyze the data from both mobile and computer simultaneously. Encase V6 software checks the data which is acquired by NEUTRINO. Neutrino Wave Shield signal blocking bag has been tested under cell towers which tests the reliability of the tool. Biggest eminence of the tool is that multiple devices connected with system can coordinate with computer evidences in a simultaneous manner. The tool acquires hardware support and parsing can be done for 75 devices which can be connected with various mobiles~\cite{numb8}.

\subsection{Android Data Recovery}
This tool is used for recovering messages, contacts, photos, and videos from android phones and tablets. It has the ability to redeem lost files and videos which occur while deleting, restoring, the files, accessing the root from memory cards inside Android devices. While recovering the data, the tool makes sure that there is no leakage of any personal information related to the user of the mobile~\cite{numb9}.

\subsection{XACT}
XACT performs   acquisition of deleted data files. The tool generally performs practical physical data acquisitions from appropriate phones or memory card and recovers deleted information. XACT enables   forensics specialists to securely dump memory of mobile phones and memory cards. It has got a unique feature to access protected and deleted content that may be critical to police investigation. While acquiring evidence, some files can be corrupt in nature which makes the evidence baseless and out of scope ~\cite{numb10}. 

\subsection{Final Mobile Forensics Tool}
It is also known as data carving tool for the forensic community. It can turn raw data into easy to understand data with a few simple clicks. It allows investigators to perform critical tasks efficiently during the investigation of cellular phones. The main feature of this tool is that it creates precise and detailed parsing of the data within file system. Once data is loaded inside the tool, it applies the appropriate parsing methods and rules to the required files and folders based on the model number of the mobile device. Results are displayed in the form of excel spreadsheets and HTML files. The tool contains multiple viewing options which include ASCII, Hex, Multi byte, Pictures, and Unicode. It has got the ability to create project workspaces when multiple devices are attached in the tool~\cite{numb11}.

\section{Challenges in Mobile Forensics}
While performing forensics, the experts should have sound knowledge about the evidence he/she extracted from a particular device. Generally, technical problems are observed by forensic specialists which are described below:

\paragraph{Operating systems and manufacturers:} All operating systems offer the same functions and options, but the way of storing data, security settings and other characteristics are different. For example: Windows mobile and Windows phone are two mobiles produced by Windows and are of different operating systems.

\paragraph{Connection Issues:}
It is noticed that most of the popular mobile forensics tools work under Windows OS only, and before connecting mobile to the computer, the former must install appropriate USB driver. In case of android OS, we can download the official driver i.e. Android SDK. The driver works with phones which are branded by Google. If the system has installed many forensic tools, then the user must be careful while using the driver packs since there might be old versions available due to which interference of the drivers can occur. Most modern phones use mini-USB/micro-USB connectors for cable connection~\cite{numb12}.

\paragraph{Hardware variations:}
A smart phone device mainly consists of microprocessor, main board, ROM and RAM memories, keyboard touch screen, radio modules, antenna, display unit, microphone and speakers, digital camera, and GPS. Operating system is generally stored in ROM and can be flashed or updated according to the requirement of the system.Depending on phone providers, manufacturers personalize the mobile in order to fulfil the demand of the market. Smartphone tends to replace quickly as compared to other devices, thus forensics examiners must have hundreds of adapters and power cords based on the type of hardware~\cite{numb13}.

\paragraph{Data volatility:}
 Many important data reside in smart phone in a volatile manner. These data might play an important role in getting strong evidence. Hence, while performing forensics, the device must be kept turned on and isolated so that important data could not get lost.

\paragraph{Anti forensic Techniques:}
Forensics becomes difficult when techniques like data hiding, data obfuscation, and data forgery are implemented on the data.

\paragraph{Generic state of the device:} Even though the device seems to be switched off, background processes may still run. For example, alarm clock in a mobile may work even if the mobile is switched off. If a sudden change in transition occurs in a device , it may result in the loss or modification of data~\cite{numb14}.

\section{Results and Conclusion}

\paragraph{}
From the survey carried out, it has been noticed that LANTERN can extract more data as compared to any other tool in the market. Also, the unique feature of NEUTRINO has been identified that it can deal with both mobile and computer simultaneously. 

\paragraph{}
Additionally, MOBILedit seems to be one of the most trusted tools for mobile phones as its generated reports are trustworthy. Despite of the sophisticated systems of IOS which keeps it protected and secure, it still manages to extract data from the iOS devices using the lockdown files method. It is commercially successful and is relied by the US Military, FBI, and CIA even in the courtrooms.

\paragraph{}
There is the presence of enough mobile forensics tools but none of them is compatible to every situation. There are certain tools which perform forensics for the iOS whereas some for Android OS and some for Windows OS.Offenders are smart enough to carry out felonious activities by integrating different operating systems and finding loopholes in the present tools. 

\paragraph{}
Therefore, we can come to a conclusion that it is necessary to keep these points in consideration while making mobile forensics tool.Finally, challenges dealt with forensics are also noticed by the offenders because of which the network security experts must have that skill to deal with those challenges manually or with some automation systems developed.   \includegraphics[width=0.1\textwidth]{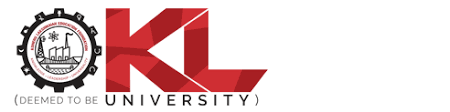}%

 \clearpage
\bibliographystyle{splncs04}
\bibliography{references}
\end{document}